\documentstyle[11pt,aaspp4]{article}
\slugcomment{accepted by Astrophysical Journal}

\begin{document}

\title{Gamma--ray Production in Supernova Remnants}

\author{T.K.~Gaisser}
\affil{Bartol Research Institute, University of Delaware, Newark, DE 19716}

\author{R.J.~Protheroe}
\affil{Department of Physics and Mathematical Physics\\
The University of Adelaide, Adelaide, Australia 5005}

\author{Todor Stanev}
\affil{Bartol Research Institute, University of Delaware, Newark, DE 19716}

\begin{abstract}
 The bulk of the cosmic rays up to about 100 TeV are thought to
 be accelerated by the 1st order Fermi mechanism at supernova
 shocks, producing a power-law spectrum.  Both electrons and
 protons should be accelerated, but their ratio on acceleration
 is not well known.  Recently, the EGRET instrument on the
 Compton Gamma Ray Observatory has observed supernova remnants IC
 443 and $\gamma$~Cygni at GeV energies.  On the assumption that
 the observed gamma-rays are produced by accelerated particles in
 the remnants (rather than, for example, from a central compact
 object) we model the contributions due to pion production,
 bremsstrahlung, and inverse Compton scattering on the cosmic
 microwave, diffuse galactic radiation, and locally produced
 radiation fields.  In the case of the same spectral index for
 both electrons and nuclei, and a cut-off at 80 TeV, we find that
 a spectral index of accelerated particles close to 2.4, and a
 ratio of electrons to protons in the range 0.2 to 0.3, gives a
 good fit to the observed spectra.  For lower cut-off energies
 flatter spectra are possible.  We also investigate the case
 where the electron spectrum is steeper than that of nuclei.  We
 discuss the implications of our results for observations at air
 shower energies, and for the propagation of cosmic rays.
\end{abstract}


\section{Introduction}

  Supernova remnants (SNR) are believed to be responsible for
 accelerating particles to energies of at least 100 TeV (see e.g.
 \cite{Drury,BlandfordEichler87,BerezhkoKrymsky88,JonesEllison91}),
 and a fraction of the accelerated particles would interact
 within the supernova remnant and produce gamma--rays
 (\cite{DruryAharonianVolk}).  Recent observations above 100 MeV
 by the EGRET instrument on the Compton Gamma Ray Observatory
 have found gamma ray signals associated with at least two
 supernova remnants -- IC~443 and $\gamma$~Cygni
 (\cite{Esposito96}).  Further evidence for acceleration in SNR
 comes from the recent ASCA observation of non-thermal X--ray
 emission from SN~1006 (\cite{Koyama95}).  Reynolds (1996) and
 Mastichiadis (1996) interpret the latter as synchrotron emission
 by electrons accelerated in the remnant up to energies as high
 as 100 TeV.

 In the diffuse galactic background the characteristic $\pi^0$
 peak at 70 MeV is clearly evident. The power-law spectra,
 dominated by bremsstrahlung, contribute $\sim 10\%$ at 1 GeV
 (\cite{Hunter95,Hunter97}).  The $\pi^0$ peak is not so clearly visible,
 however, in the spectra of IC~443 and $\gamma$~Cygni, possibly
 due to the larger error bars.  However, this may instead suggest
 larger bremsstrahlung and inverse Compton (IC) contributions,
 and a correspondingly higher electron to proton ratio than in
 galactic cosmic rays.  From detailed modeling of the gamma-ray
 spectra of IC~443 and $\gamma$~Cygni it should be possible to
 determine the relative contributions of $\pi^0$ decay,
 bremsstrahlung, and inverse Compton scattering to these spectra,
 and thereby to obtain information about the conditions in these
 sources. This would also allow a better determination
 of the expected extrapolation of the Egret
 spectra for these sources to the TeV energy range.

 Gamma rays are produced by electrons in bremsstrahlung
 interactions, and by inverse Compton interactions with the
 cosmic microwave background (\cite{Mastichiadis96}), with the
 diffuse galactic infrared/optical radiation, and with the
 radiation fields of the remnant itself.  Protons produce gamma
 rays through the decay of neutral pions produced in
 proton--nucleus collisions.  Drury, Aharonian \& V\"{o}lk (1993)
 calculated the $\pi^0$ decay gamma ray flux expected from
 supernova remnants due to interactions of accelerated cosmic ray
 nuclei with matter in the remnant.  They calculated the expected
 flux as a function of supernova age and showed that a number of
 remnants should be detectable above 100 MeV.  Mastichiadis
 (1996) has included synchrotron radiation and inverse Compton
 scattering on the cosmic microwave background, and shown that
 synchrotron radiation by directly accelerated electrons is
 probably responsible for the non-thermal X--rays observed in
 SN~1006.  Mastichiadis also showed that SNR could produce a flux
 at TeV energies by inverse Compton scattering which is
 comparable to that predicted for pion production by Drury,
 Aharonian \& V\"{o}lk (1993) .  An important point of the papers
 of both Reynolds (1996) and Mastichiadis (1996) is that
 electrons should be accelerated up to 100 TeV energies.
 Mastichiadis also suggests that the observation of TeV gamma
 rays from supernova remnants would not, in itself, provide
 direct evidence of acceleration of nuclei at supernova shocks.
 Pohl (1996) has also recently suggested TeV gamma-rays from
 supernova remnants may be of leptonic origin.

 In the present paper we consider gamma-ray production by
 interactions of accelerated nuclei and electrons in IC~443 and
 $\gamma$~Cygni.  Our approach is to analyze the observed
 $\gamma$--ray fluxes and attempt to extract the source
 parameters from the data rather than use theoretical models of
 particle acceleration in supernova remnants. 
 We assume power law  spectra for electrons and protons, possibly
 different in slope and in maximum momentum on acceleration.
 We perform a
 maximum likelihood fit to the gamma ray spectra of IC~443 and
 $\gamma$~Cygni to determine the ratio of electrons to protons,
 the power-law spectral index, and the average matter density
 seen by the accelerated particles.

 \section{Relative importance of pion production, bremsstrahlung
 and inverse Compton scattering}

 Before describing our detailed calculation, we first briefly
 discuss the general features of the expected $\pi^0$,
 bremsstrahlung and inverse Compton gamma ray spectra, and make
 order of magnitude estimates of their relative contributions to
 the gamma ray flux.  The momentum spectrum of particles
 accelerated by first order Fermi acceleration at a strong shock
 is expected to be $dN/dp \sim p^{-2}$.  Protons accelerated with
 this spectrum, and interacting within the remnant, will produce
 at high energies a $\pi^0$ gamma ray number spectrum having
 approximately the same slope, $\sim E^{-2}$, but with a
 low-energy turnover at $\sim m_\pi c^2/2$.  At high energy,
 electrons will cool by inverse Compton and synchrotron losses,
 $dE/dt \propto -E^2$ (neglecting, for the moment, scattering in
 the Klein-Nishina regime).  In an equilibrium situation, the
 ambient spectrum of electrons will be steepened by one power to
 $\sim E^{-3}$, giving rise to a spectrum of inverse Compton
 gamma rays proportional to $\sim E^{-2}$.  However, as we shall
 show below, {\it a priori} it is not obvious that that such an
 equilibrium exists for the supernova remnants in question.

 The cooling time for electrons in a magnetic field is
\begin{equation} t_{\rm syn} \approx 1.3 \times 10^{10} \left({B
 \over 1 \; \mu {\rm G}} \right)^{-2} \left({E \over 1 \; {\rm
 GeV}} \right)^{-1} \hspace{5mm} {\rm years}.  
\end{equation} 
 So, for typical interstellar magnetic fields ($\sim 3 \, \mu$G),
 one finds a cooling time at 100 TeV of $\sim 1.4 \times 10^4$
 years.  This time is longer than, but of the same order of
 magnitude of the age of the supernova remnants we are
 considering.  Because of this, Mastichiadis (1996) assumed
 that the ambient electron spectrum
 is not yet affected by radiative losses.
 The cooling time for IC scattering on the microwave background
 is of the same magnitude and so IC losses will have some effect,
 though possibly minor, on the shape of the electron spectrum.
 
 Hence, as a starting point we estimate the
 $\gamma$--ray luminosities assuming that protons and electrons
 are accelerated with identical power law momentum spectra
 extending to the same maximum momentum $p^{\rm max}$, and that these
 spectra have not significantly evolved.  The bremsstrahlung
 spectrum will have the same power-law as the electron spectrum,
 extending to low energy so that bremsstrahlung photons
 eventually dominate below the $\pi^0$ peak even for very small
 $e/p$ ratios.  On the other hand, the inverse Compton spectrum
 from an $E^{-2}$ electron spectrum will be $\sim E^{-3/2}$,
 which is much flatter than the $\pi^0$ and bremsstrahlung gamma
 ray spectra.

 We next discuss the relative importance of the various processes
 to the gamma ray flux based on order of magnitude estimates.
 The crude approximations discussed below are only used for these
 order of magnitude estimates, and we use exact formulae for the
 results we shall present later.  Furthermore, in section 5 we
 shall make detailed fits to the gamma ray spectra which allow
 for the possibility of the ambient electron spectrum being
 steeper than the proton spectrum, either due to a different
 spectrum at acceleration or due to evolution of the electron
 spectrum.

 For this estimate we approximate the total proton spectrum (integrated
 over the SNR) by $N(E_p) \equiv dN/dE_p = a_p E_p^{-\alpha}$
 protons~GeV$^{-1}$.  For $E_\gamma\gg m_\pi c^2/2$, the gamma-ray
 luminosity from $\pi^0$ production is
\begin{equation}
L_{\pi^0}(E_\gamma) 
\approx c\,n\,\left(\sigma_{pp}^{inel}\,{2Z_{N\pi^0}\over\alpha}\right)
\times a_p\,E_\gamma^{-\alpha},
\end{equation}
 where $\sigma_{pp}^{inel}$ is the inelastic proton--proton cross
 section, $\rho \approx n\,m_p$ is the average matter density
 sampled by the protons and $Z_{N\pi}$ is a spectrum-weighted
 moment of the momentum distribution of pions produced in
 proton-proton collisions (\cite{Gaisser}).  For $\alpha =$~2.0,
 2.3, and 2.4 respectively, $Z_{N\pi^0} \approx$ 0.16, 0.075 and
 0.066.  Thus, for $\alpha = 2.0$
\begin{equation}
L_{\pi^0}(E_\gamma)\approx 1.5\times 10^{-16}\,a_p\,n\,E_\gamma^{-2}
 \hspace{5mm} {\rm photons \; GeV^{-1} \; s^{-1}}
\end{equation}
 where $E_\gamma$ is in GeV and $n$ is in cm$^{-3}$.

 Similarly, we approximate the total electron spectrum by $N(E_e)
 \equiv dN/dE_e = a_e E_e^{-\alpha}$ electrons GeV$^{-1}$.  To
 obtain the bremsstrahlung luminosity, we assume that after an
 electron of energy $E_e$ has traveled one radiation length,
 $X_0$, it is converted into a photon of energy $E_\gamma = E_e$.
 Hence,
\begin{equation} 
L_{\rm brem}(E_\gamma) \approx N(E_\gamma) \rho c / X_0.
\end{equation} 
 Thus, for $\alpha = 2.0$, 
\begin{equation} L_{\rm
 brem}(E_\gamma) \approx 7 \times 10^{-16} a_e n E_\gamma^{-2}
 \hspace{5mm} {\rm photons \; GeV^{-1} \; s^{-1}}.
\end{equation} 

 For inverse Compton scattering, we approximate the photon energy
 after scattering by an electron of energy $E_e = \gamma m_ec^2$
 by $\gamma^2 \bar{\varepsilon}$ where $\bar{\varepsilon}$ is the mean 
 photon energy of the radiation field under consideration.
 Provided the Compton scattering is in the Thomson regime 
 ($\gamma \bar{\varepsilon} \ll m_e c^2$) this gives an inverse Compton
 luminosity
 \begin{equation}
 L_{\rm IC}(E_\gamma) \approx N(\gamma) n_{\rm ph} \sigma_T c / 2 \gamma 
 \bar{\varepsilon}
 \end{equation}
 where $N(\gamma) d\gamma = N(E_e) d E_e$,  and we obtain
 \begin{equation}
 L_{\rm IC}(E_\gamma) \approx a_e 
 {(\bar{\varepsilon})^{1/2} \over E_\gamma^{3/2}}
 {n_{\rm ph} \sigma_T c \over m_e c^2}.
 \label{Eq:approxIC}
 \end{equation}
 For scattering off the microwave background, we use 
 $n_{\rm ph} = 400$ cm$^{-3}$, $\bar{\varepsilon} = 6.25 \times 10 ^{-4}$ eV,
 and obtain
 \begin{equation}
 L_{\rm IC}(E_\gamma) \approx 1.3 \times 10^{-14} a_e E_\gamma^{-3/2} 
 \hspace{5mm} {\rm photons \; GeV^{-1} \; s^{-1}}
\end{equation}
 where $E_\gamma$ is in GeV.  We note that, for an assumed matter
 density of 1 cm$^{-3}$, at 1 GeV the inverse Compton scattering
 contribution is an order of magnitude larger than the
 bremsstrahlung contribution, and the relative importance of the
 inverse Compton scattering contribution increases with energy.
 In the next two sections we describe the radiation and matter
 environments of the two SNR, and our accurate treatment of pion
 production, bremsstrahlung, and inverse Compton scattering.

 \section{The environments of IC~443 and $\gamma$~Cygni}

 In addition to the microwave background radiation, and the galactic
 infrared/optical background radiation, the radiation
 produced locally within the SNR will provide target photons 
 for inverse Compton scattering.
 The two SNR we consider are at galactocentric radii sufficiently
 close to that of the Sun that we may use the local infrared/optical
 radiation field for the galactic background.
 We adopt the spectrum of Mathis et al. (1983)
 which is the sum of 6 diluted blackbody spectra and is
 shown in Figure~1.
 The mean photon energy for this spectrum is 
 $\bar{\varepsilon} \approx 0.08$ eV, and
 the total photon number density is $n_{\rm ph} \approx 8.6$ cm$^{-3}$,
 giving an energy density of $U_{\rm ph} \approx 0.66$ eV cm$^{-3}$.
 Using Equation~\ref{Eq:approxIC} we can estimate 
 the gamma ray luminosity applicable to the Thomson regime 
 \begin{equation}
 L_{\rm IC}(E_\gamma) \approx 3.1 \times 10^{-15} a_e E_\gamma^{-3/2} 
 \hspace{5mm} {\rm photons \; GeV^{-1} \; s^{-1}.}
 \end{equation}
 This is only a factor $\sim 4$ lower than that for scattering off
 the microwave background, and so it is necessary to include this 
 radiation field in an accurate calculation.

 Lozinskaya (1991) gives an excellent discussion of the
 supernova remnants IC~443 and $\gamma$~Cygni.
 We consider first the remnant IC~443 which is at a distance of 
 $\sim 1.5$ kpc, and has an angular diameter of 
 $\sim 45^\prime$ giving a radius of $\sim 10$ pc.
 The remnant is probably about 5000 years old and is expanding into 
 a very inhomogeneous medium including dense molecular clouds.
 Most of the remnant has a temperature of $\sim 1.2 \times 10^7$ K
 (\cite{Petre88}).
 Recent measurements by Asaoka and Aschenbach (1994)
 of the X--ray spectrum indicate two components:
 $\sim 27$ M$_\odot$ at 1 KeV, and $\sim 6$ M$_\odot$ at 13 KeV.
 Identifying the swept-up mass with $27$ M$_\odot$ leads to a 
 pre-SNR interstellar density of $\sim 0.3$ cm$^{-3}$.
 Petre et al. (1988) measured the X--ray flux in the
 2--10 keV range, $6.7 \times 10^{-11}$ erg cm$^{-2}$ s$^{-1}$,
 and a temperature of 1.03 keV.
 Approximating the spectrum by a thermal bremsstrahlung spectrum,
 we obtain a total X--ray energy flux of $3 \times 10^2$ 
 eV cm$^{-2}$ s$^{-1}$.
 Dividing by the solid angle subtended by the SNR,
 and multiplying by $4 \pi/c$ we get an approximation to the 
 X--ray energy density in the remnant of 
 $3 \times 10^{-3}$ eV cm$^{-3}$.
 For a mean photon energy of $\bar{\varepsilon} \approx  1$ keV,
 one obtains a photon number density of 
 $n_{\rm ph} \approx 10^{-6}$ cm$^{-3}$.
 Using Equation~\ref{Eq:approxIC} we can estimate the gamma ray luminosity
 applicable to the Thomson regime (in this case for $E_\gamma \ll 1$ GeV)
 \begin{equation}
 L_{\rm IC}(E_\gamma) \approx 4.2 \times 10^{-20} a_e E_\gamma^{-3/2} 
 \hspace{5mm} {\rm photons \; GeV^{-1} \; s^{-1}.}
 \end{equation}
 At higher energies (where scattering is in the Klein-Nishina regime)
 the gamma ray luminosity will be much lower.
 Comparing the flux from inverse Compton scattering on the X--ray 
 emission with that from inverse Compton scattering on the microwave
 background, we find it to be negligible in IC~443.
 However, in younger supernova remnants the X--ray emission may present
 a significant target for IC scattering.

 The supernova remnant associated with $\gamma$~Cygni, G78.2+2.1,
 is at a distance of $\sim 1.8$ kpc, and has an angular size of 
 $\sim 1^\circ$ giving a radius of $\sim 16$ pc.
 The remnant is probably about 7000 years old and is also expanding into 
 a very inhomogeneous medium including dense molecular clouds.
 A dense cloud occupies $\sim 5$\% of the volume of the remnant
 and this has been used by Pollock (1985)
 to predict the gamma ray flux
 (see also \cite{Aharonian94}).
 We shall return to a discussion of the effects of an inhomogeneous
 interstellar medium later when we discuss the electron to
 proton ratio.
 The X--ray flux obtained by Higgs, Landecker, \& Seward (1983) in the
 0.2--4 keV range is about $6 \times 10^{-11}$ erg cm$^{-2}$ s$^{-1}$,
 with a temperature of 1.3 keV and ambient gas density of 
 $\sim 0.2$ cm$^{-3}$.
 The X-ray intensity is of the same order of magnitude as from IC 443, and so 
 we conclude that inverse Compton scattering on the X--rays in 
 $\gamma$~Cygni,
 and other old SNR, will not make an important contribution to the 
 gamma ray flux.

 A major component of the radiation field of SNR is the infrared emission
 due to shock heated dust.
 For both sources, Saken, Fesen, \& Shull (1992) provide infrared spectra
 obtained with the IRAS satellite.
 We use their two-temperature model fits to represent the radiation
 field in the infrared.
 We have used the solid angle subtended by each source 
 to obtain the energy density from the intensity.
 In IC~443 the mean photon energies, number densities, and 
 energy densities of the two diluted blackbody components are: 
  $\bar{\varepsilon} \approx 0.008$, 0.04 eV; 
 $n_{\rm ph} \approx 18.4$, 11.8 cm$^{-3}$;
 $U_{\rm ph} \approx 0.15$, 0.51 eV cm$^{-3}$.
 In $\gamma$~Cygni the mean photon energies, number densities, and 
 energy densities of the two diluted blackbody components are: 
 $\bar{\varepsilon} \approx 0.009$, 0.025 eV; 
 $n_{\rm ph} \approx 37.3$, 6.2 cm$^{-3}$;
 $U_{\rm ph} \approx 0.32$, 0.15 eV cm$^{-3}$.
 In both cases the infrared photon densities exceed the diffuse
 Galactic infrared photon densities, and so inverse Compton scattering
 of locally produced infrared photons could make a substantial
 contribution to the total inverse Compton flux.
 The total radiation field we adopt is shown in Fig.~2(a) for IC~443 
 and in Fig.~2(b) for $\gamma$~Cygni, and in each case we show 
 the contributions of the diffuse background and the locally 
 produced radiation.  The IRAS data (\cite{Saken92,Mufson86}) are also shown.

 \section{Accurate treatment of pion production, 
 bremsstrahlung and inverse Compton scattering}

  For interactions with matter we assume standard interstellar
 composition. For proton interactions we use the event generator
 TARGET (Gaisser, Protheroe \& Stanev 1983) 
 in its proton target version. TARGET, which has
 been extensively tested in numerous applications, represents correctly
 the proton interaction cross sections and the secondary particles 
 spectra starting at the pion production production threshold. At
 energies above 100 GeV the interaction cross section has
 a $\ln^2{s}$ behavior for proton--proton interactions. At total
 proton energy below 2.5 GeV TARGET generates only $\Delta$ resonances.
 The output of this part of the code was compared with the results of
 Dermer (1986) and is in good agreement. 

 In the case of bremsstrahlung, 
 yields are calculated for fully ionized matter for which the 
 cross--section is strongly energy dependent. 
 We use the expressions of Koch \& Motz (1959)
 with form factors for hydrogen and helium adjusted to 
 represent the more precise values of Tsai (1974).
 The cross-sections for fully ionized matter are calculated
 with the bremsstrahlung formulae valid in the absence of
 screening by the atomic electrons.
 These cross sections are given by Protheroe, Stanev, \& Berezinsky (1995).

 For inverse Compton scattering we use the Klein-Nishina cross section
 (\cite{Jauch}) to calculate the mean free path and the distribution of
 photon energies produced in the IC process. We perform an exact Monte
 Carlo simulation of interactions as described by  Protheroe, Mastichiadis
 \& Dermer (1992) to build up distributions of secondary electrons and
 photons arising from interactions of electrons of a given energy with
 blackbody radiation of given temperature. These distributions are then
 convoluted with the parent electron spectrum.

 \section{Results and comparison to experimental data}

 We assume the momentum spectrum of accelerated particles is of the form
 \begin{equation}
 {d N \over d p} = ac \left( {p \over {\rm 1 \, GeV/c}} \right)^{-\alpha} 
 \hspace{5mm} {\rm (GeV/c)^{-1}}
 \label{Eq:spectrum}
 \end{equation}
 so that the energy spectrum is
 \begin{equation}
 {d N \over d E} = a \left( {E \over {\rm 1 \,GeV}} \right) 
 \left( {p \over {\rm 1 \, GeV/c}} \right)^{-(1+\alpha)}
 \hspace{5mm} {\rm GeV^{-1}}
 \end{equation}
 where $E=(p^2c^2 + m^2c^4)^{1/2}$.  
 We allow for different normalizations and spectral indices of
 protons ($a_p$, $\alpha_p$) and electrons ($a_e$, $\alpha_e$),
 and define the electron to proton ratio as  $R_e \equiv a_e/a_p$
 which is the ratio of the differential momentum spectra at 1~GeV/c.
 At high energies ($E \gg m_p c^2$) the energy spectrum of
 protons is just $d N_p / d E \approx a_p E^{-\alpha_p}$
 GeV$^{-1}$, which gives the total number of protons per unit
 energy at the supernova remnant.  Similarly, for electrons at
 high energies ($E \gg m_e c^2$) the energy spectrum is $d N_e /
 d E \approx a_e E^{-\alpha_e}$ GeV$^{-1}$.

 We calculate gamma ray emissivities for $\pi^0$ production, 
 bremsstrahlung, and inverse Compton scattering in the radiation 
 fields appropriate to each SNR for the case where 
 $a_p/V = 1$ GeV$^{-1}$ cm$^{-3}$ ($V$ is the volume of the
 region where the accelerated particles are located), $R_e = 1$,
 and a nucleon density of $n=1$ cm$^{-3}$.
 The emissivities are calculated for each process for a range of $\alpha$
 for the case where $\alpha_e =\alpha_p = \alpha$, 
 assuming $n=1$ cm$^{-3}$, $R_e=1$, and a radiation field
 analogous to Fig.~2a (IC~443).  The emissivity, $Q(E_\gamma)
 \equiv dQ_\gamma/dE_\gamma$ (cm$^{-3}$ s$^{-1}$ GeV$^{-1}$),
 multiplied by $E_\gamma^2$, is shown in Fig.~3 for $\alpha=2$.
 With the exception of the local infrared radiation field (short
 dashes in Fig.~3), and neglecting the small differences of the
 galactic infrared background as a function of the galactocentric
 distance, this emissivity spectrum is universal for SNR in the
 Galaxy (with suitable scaling of $\pi^0$ and bremsstrahlung
 emissivity with matter density) on the assumption that supernova
 remnants accelerate charged particles with a $p^{-2}$ spectrum,
 and that radiative losses have not yet affected the electron
 spectrum. The main feature of the emissivity spectrum is the
 dominance of the inverse Compton scattering (as already expected
 from Eq.~\ref{Eq:approxIC}) with its very flat spectrum.

  It is instructive to follow the scaling of the emissivities of
 different processes with the supernova remnant parameters.
 The relative contribution of IC scales up or down only with the
 electron to proton ratio. In the vicinity of $E_\gamma$ = 1 GeV
 the electron to proton ratio has to be of order of 0.01 for the
 IC scattering on the microwave background not to exceed the
 $\pi^0$ contribution for a matter density of 1 nucleon cm$^{-3}$
 and $\alpha$ = 2. As long as IC is the most important contribution,
 the gamma-ray spectrum will be harder than the parent electron spectrum.
 The $\pi^0$ contribution scales only with the matter density.
 Alternatively, if the matter density is very high, 
 $\sim 100$ nucleons cm$^{-3}$ one can suppress the IC contribution
 relative to the bremsstrahlung and $\pi^0$ contributions. 
 For such high densities, more than two orders of
 magnitude higher than the average densities expected from the
 mass of the ejecta (\cite{Lozinskaya}), the contribution  of
 bremsstrahlung, which is proportional both to matter density and to
 the electron to proton ratio, would dominate. For a $\pi^0$ `bump'
 to be visible requires an electron to proton ratio less than 1.

We define $Q_0^\pi$, $Q_0^{\rm brem}$, and $Q_0^{\rm IC}$ to be the 
gamma ray emissivities (cm$^{-3}$ s$^{-1}$ GeV$^{-1}$)
for pion production, bremsstrahlung, and inverse Compton scattering
calculated for $A \equiv a_p/V = 1$ GeV$^{-1}$ cm$^{-3}$, $R_e=1$, 
and $n=1$ cm$^{-3}$.
Then the gamma ray flux observed at Earth, a distance $d$ from the SNR, 
is given by
\begin{equation}
F_\gamma(E_\gamma,\alpha) = {n_1 A_1 V \over 4 \pi d^2} 
\left[ Q_0^\pi (E_\gamma,\alpha) + R_e Q_0^{\rm brem} (E_\gamma,\alpha) + 
{ R_e \over n_1} Q_0^{\rm IC} (E_\gamma,\alpha) \right]
\end{equation}
where $n_1 \equiv n/(1$~cm$^{-3}$), 
$n$ is the nucleon number density in the 
region of the SNR containing the accelerated particles, and
$A_1 \equiv A/(1$~GeV$^{-1}$ cm$^{-3}$).
We perform a maximum likelihood fit to the EGRET data
(\cite{Esposito96}) for each SNR with the following free
parameters: electron to proton ratio $R_e$, particle spectral
indices ($\alpha_p$, $\alpha_e$), maximum energy $E^{\rm max}$
(corresponding to an exponential cutoff,
i.e. $\exp(-E/E^{\rm max})$), which could also be different for
protons and electrons, $R_e/n_1$, and the overall normalization
\begin{equation}
B \equiv {n_1 A_1 V  \over 4 \pi d^2} 
= \left( {a_p \over 1 \; {\rm GeV^{-1} \; cm^{-3}}} \right) 
{n_1 \over 4 \pi d^2} 
\end{equation}
which has units cm GeV$^{-1}$.

 For each SNR, the $\gamma$--rays are binned in 10 energy bins,
 two of which give upper limits, and so the total number of data
 points is not sufficiently large to fit all possible parameters
 simultaneously.  Instead we perform four different fits
 corresponding to the following assumptions.

  Fit~1. Both protons and electrons are accelerated with
 power-law spectra with the same index $\alpha$ and to the same
 maximum energy of 80 TeV. The choice of $E^{\rm max}$=80 TeV is
 arbitrary but sufficiently high for fitting the EGRET
 $\gamma$--ray spectra which only extend to $\sim$10 GeV and
 for extrapolating to the Whipple limits of several hundred GeV.
 The assumption of a pure power law is an approximation.
 Non-linear effects in cosmic-ray modified shocks generally
 lead to some concavity in the spectra (\cite{JonesEllison91,Berezhko96}),
 though the deviation from a power-law is slight for relativistic
 particles.  For example, the proton spectrum calculated by Ellison
 (1993) steepens by less than 2 per cent in the spectral
 index below 1 TeV from its flattest high energy value.  

  Fit~2. Protons and electrons are accelerated with the same
 spectral index $\alpha$ but the electron spectrum cuts off at an
 energy $E_e^{\rm max}$ which is a fit parameter. $E_p^{\rm max}$
 = 80 TeV.

  Fit~3. Protons and electrons are accelerated to different
 power law spectra with indices $\alpha_p$ and $\alpha_e$. The
 maximum acceleration energy for electrons is 1000 GeV.
 $E_p^{\rm max}$ = 80 TeV.

  Fit~4. Protons and electrons are accelerated to identical power
 law spectra and both cut off at the same maximum energy $E^{\rm
 max}$.

  The fitting procedure actually consisted of a tabulation of the
 maximum likelihood value $L$ for a pre--determined set of
 parameter values. Except for the spectral indices $\alpha$ which
 were taken on a 0.05 grid, all other parameters were introduced
 on logarithmic scales of 10$^{0.1}$, i.e. ten parameter values
 per decade. This grid for the trial parameters is small enough
 to produce neighboring spectra well within the uncertainties of
 the measurements. The parameter set with a maximum $L$ value
 was taken to be the best fit.

\subsection{Results from Fit~1}

 Fig.~4 shows the fitted spectrum for the supernova remnant IC~443 in Fit~1.
 The upper limits on the VHE $\gamma$ rays from IC~443 obtained by the
 Whipple observatory (\cite{Buckley97}) and from the HEGRA
 array (\cite{Prosch95}) are also shown in Fig.~4, and seen to be
 consistent with the fit, even though they were not used in the fit.
   
 The fit gives $\alpha$ = 2.32$\pm^{0.14}_{0.11}$. The errors are
 calculated from the values of $\alpha$ at which $\ln L = \ln L_0
 - s^2/2$, where $L_0$ is the best fit value of the likelihood
 function $L$, and $s$ is the number of standard deviations
 ($s=1$ to give $1\sigma$ errors).  The overall normalization
 gives $B = 4.0 \times 10^8$ cm GeV$^{-1}$.  The electron to
 proton ratio corresponding to this value of $\alpha$ is
 $R_e=0.16\pm^{0.29}_{0.08}$.  The best fit value of $R_e/n_1$ is
 $\sim 10^{-4}$, with very large error bars because the emission
 is dominated by $\pi^0$ production and bremsstrahlung.  Thus, we
 can not reliably give the matter density since the fit requires
 a very high density only to make the IC component negligible.
 This also prevents us from obtaining a reliable estimate of the
 cosmic ray density at the source.

  Fig.~5 shows a similar fit for $\gamma$~Cygni. The best $\alpha$
 value is $\alpha = 2.42\pm^{0.09}_{0.07}$, which gives an
 electron to proton ratio of $R_e = 0.16\pm^{0.14}_{0.08}$,
 and a similar value of $R_e/n_1$ to IC~443.
 The overall normalization gives $B = 10^9$ cm GeV$^{-1}$.
 The VHE limits are from Whipple (\cite{Buckley97}) and from 
 HEGRA (\cite{Prosch96,Willmer95}) and are again 
 consistent with the fit.

 We can get some useful information from the overall
normalization constant $B \equiv n_1 A_1 V / 4 \pi d^2$ which,
given that we know $d$, gives us the product of $n$,
$N_p(p=1$~GeV/c$)/V$, and $V$.  The total energy content of a
spectrum of the form of Equation~\ref{Eq:spectrum} with
$\alpha=2.32$ and $a_p=1$ is $\sim 3$~GeV.  Thus for IC~443 at a
distance of 1.5~kpc the normalization $B=4.0\times 10^8$ cm
GeV$^{-1}$ requires a total energy in accelerated protons of
\begin{equation}
U_{\rm CR} \sim 5 \times 10^{50} 
\left( {n \over {\rm 1 \; cm^{-3}}} \right)^{-1}
\hspace{5mm} {\rm erg},
\label{Eq:energy}
\end{equation}
where $n$ is the number density of gas in the region where the
accelerated particles are located.  The corresponding value for
$\gamma$~Cygni is a factor of 3 higher.  Alternatively, we can
obtain an estimate of the total mass in the region where the
accelerated particles are located as a function of the energy
density in accelerated particles
\begin{equation}
M \sim 3 \times 10^5 
\left( {u_{\rm CR} \over {\rm 1 \; eV  \; cm^{-3}}} \right)^{-1}
\hspace{5mm} {\rm M_\odot}
\end{equation}
for IC~443 and a factor 3 higher for $\gamma$~Cygni.

 Fit~1 requires a very high matter density in order to suppress
 the IC contribution to the $\gamma$--ray flux which is much
 flatter than the observed spectrum.  The SNR G78.2+2.1
 associated with $\gamma$~Cygni was tentatively identified with
 the COS-B source 2CG78+01 (\cite{Pollock85A}).  A cloud of
 density $\sim 300$ cm$^{-3}$ occupies $\sim 5$\% of the volume
 of the remnant, and this density has been used by Pollock (1985)
 in predicting GeV gamma ray fluxes.  Aharonian, Drury, \&
 V\"{o}lk (1994) also discussed this association and point out
 that the emission may extend to TeV energies.  At radio
 frequencies, most of the emission comes from the SE part of the
 remnant (\cite{Higgs83A}), and this appears to coincide also
 with a region of exceptionally high 90 $\mu$m intensity, and a
 molecular cloud near the rim of the remnant, implying that this
 region may be the source of the gamma ray emission.  A density
 of $\sim 300$ cm$^{-3}$ is comparable with the densities that
 come from the maximum likelihood fits.  However, the fitted
 density is the density averaged over the region where the
 accelerated particles are located, and so this value may be
 unrealistically high unless acceleration is taking place only
 close to such a massive high-density cloud.  This might occur at
 massive clouds interacting with the SNR shock.

  Another way to look at the numbers for $\gamma$~Cygni is to
 note that the energy in cosmic rays required in a region with
 density $300$~cm$^{-3}$ is $5\times 10^{48}$~erg.  If this high
 density occupies 5\% of the volume of the SNR then the total
 energy in cosmic rays, assuming them to be uniformly
 distributed, is $\sim 10^{50}$~erg, approximately 10\% of the
 initial kinetic energy of the ejecta of a typical supernova.


Finally, we check that for the fitted electron spectra the
synchrotron X--ray flux predicted for reasonable magnetic fields
does not exceed that observed.  This is particularly important
for the case of Fit~1 where the electron spectrum extends up to
80 TeV and one would expect to generate a significant X--ray flux
for any standard value of magnetic field at the shock.

Taking the electron spectrum in the source to be 
\begin{equation}
N(E) \equiv {dN \over dE} = a_e E^{-\alpha} \exp(-E/E_e^{\rm max}) 
\; \; \; \; \rm GeV^{-1},
\end{equation}
where 
$E$ is in GeV, we may obtain the synchrotron X--ray flux at
Earth (erg cm$^{-2}$) in the energy range $\epsilon_1$ to $\epsilon_2$
\begin{equation}
F_X = {a_e \over 4 \pi d^2} \int_{\epsilon_1/h}^{\epsilon_2/h} d \nu
\int dE E^{-\alpha} P(\nu, E, H_\perp).
\end{equation}
Here $P(\nu, E, H_\perp)$ is the power per unit frequency (erg Hz$^{-1}$)
emitted at frequency $\nu$ by an electron of energy $E$ (GeV) in a
magnetic field with perpendicular component $H_\perp$ (gauss) 
and is given by
\begin{equation}
P(\nu, E, H_\perp) = 4 \pi c_3 H_\perp F(x)
\end{equation}
where $c_3= 1.87 \times 10^{-23}$, $x=\nu/\nu_c$, $\nu_c \approx
c_1 H_\perp (625 E)^2$, $c_1=6.27 \times 10^{18}$, and
\begin{equation}
F(x) = x \int_x^\infty K_{5/3}(z) dz
\end{equation}
where $K_{5/3}$ is a Bessel function of the second kind
(\cite {Pacholczyk70}).
>From Eq.~14 we see that
\begin{equation}
a_e = {B R_e \over n} 4 \pi d^2,
\end{equation}
and so we obtain 
\begin{equation} 
F_X = 4 \pi c_3 H_\perp {B R_e \over n}
\int_{\epsilon_1/h}^{\epsilon_2/h} d \nu
\int dE F(x) E^{-\alpha}  \exp(-E/E_e^{\rm max}).
\end{equation}

For the best fit parameters given above, and a perpendicular
component of magnetic field of $H_\perp =6 \mu$G, we predict for
IC~443 a 2---10~keV flux of $6.8 \times 10^{-13}$ erg cm$^{-2}$
s$^{-1}$ which is not in conflict with the observed flux of $6.7
\times 10^{-11}$ erg cm$^{-2}$ s$^{-1}$.  Similarly for
$\gamma$~Cygni we predict a 0.2---4~keV flux of $8.2 \times
10^{-12}$ erg cm$^{-2}$ s$^{-1}$ which is not in conflict with
the observed flux of $6 \times 10^{-11}$ erg cm$^{-2}$ s$^{-1}$.

 \subsection {Results from Fits~2, 3, and 4}

 As noted above, the X-ray luminosity calculated using a magnetic
 field of 6 $\mu$G and the electron spectra from Fit~1 is not in
 direct contradiction with the observed luminosities of either
 object.  However, for higher magnetic fields, e.g. due to shock
 compression, the predicted luminosity would exceed that observed
 if $H_\perp$ is higher than $\sim 54 \mu$G (IC~443) or $\sim 42 \mu$G
 ($\gamma$~Cygni).  Even for fields below these limits, there may
 be a problem due to the power-law nature of the synchrotron
 spectrum given that the observed X--ray spectrum appears to be
 consistent with thermal bremsstrahlung origin.

 We therefore perform Fits~2, 3, and 4, which tend to suppress the
 X--ray production either by cutting off the electron spectrum at
 an arbitrary $E_e^{\rm max}$ or by allowing a steeper electron
 spectrum. Tables I and II list the results of all four fits for
 the two supernova remnants. Columns 2 \& 3 give the proton and
 electron acceleration spectra, column 4 -- the cutoff energy for
 electrons (and protons in Fit~4), column 5 contains the
 preferred $R_e$ value, column 6 -- the matter density. Column 7
 gives the measure of the fit improvement ($\ln{(L/L_0)}$) and
 column 8 -- the maximum value of the magnetic field allowed by
 the observed X--ray luminosity of the source. Note that the
 synchrotron X--ray flux does not scale simply as a power-law in
 magnetic field (with exponent related the electron spectral index)
 because the X--ray emitting electrons are near or above $E_e^{\rm max}$
 where the spectrum is affected by the exponential cut-off so that the
 delta-function approximation to $F(x)$ which is often used breaks down.

\begin{table*}
\tablenum{1}
\caption{ Fit parameters for IC 443.  $E^{\rm max}_p$ fixed at 80 TeV for
 fits 1, 2 \& 3. $E^{\rm max}_e$ fixed at 80 TeV for fit 1 and at 1 TeV
 in fit 3.
 \label{tbl1}}
\begin{center}
\begin{tabular}{rccccccr}
\tableline
 Fit & $\alpha_p$ & $\alpha_e$ & $E_e^{\rm max}$ & $R_e$ & $n$
     & ln$(L/L_0)$ & $H_\perp^{\rm max}$ \\
     &    &   & (GeV) &   & (cm$^{-3}$) & & (G) \\ 
\tableline
1 & 2.32$^{+0.14}_{-0.11}$ & $=\alpha_p$ & $8 \times 10^4$ &
 0.16$^{+0.29}_{-0.08}$ & $>10^3$ & 0    &  $5.4 \times 10^{-5}$ \\
2 & 2.15$^{+0.15}_{-0.15}$ & $=\alpha_p$ & 25$^{+15}_{-10}$&
 0.4$^{+2.0}_{-0.3}$ & 0.6$^{+0.4}_{-0.2}$ & 0.36 &  $7.0 \times 10^{-2}$ \\
3 & 2.25$^{+0.20}_{-0.20}$ & 2.70$^{+0.15}_{-0.35}$ & 10$^3$ &
 0.05$^{+0.15}_{-0.02}$ & 6.3$^{+10.}_{-5.0}$ & 0.68 &  $5.3 \times 10^{-4}$ \\
4 & 1.85$^{+0.30}_{-0.15}$ & $=\alpha_p$ & 40$^{+23}_{-24}$&
 0.03$^{+0.07}_{-0.02}$ & 0.08$^{+0.12}_{-0.02}$ & 1.22 & 
 $1.2 \times 10^{-2}$ \\
\tableline
\end{tabular}
\end{center}
\end{table*}
\begin{table*}
\tablenum{2}
\caption{ Fit parameters for $\gamma$~Cygni. $E^{\rm max}_p$ fixed at
  80 TeV for fits 1, 2 \& 3. $E^{\rm max}_e$ fixed at 80 TeV for fit 1
  and at 1 TeV in fit 3.
  \label{tbl2}}
\begin{center}
\begin{tabular}{rccccccr}
\tableline
 Fit & $\alpha_p$ & $\alpha_e$ & $E_e^{\rm max}$ & $R_e$ & $n$
     & ln$(L/L_0)$ & $H_\perp^{\rm max}$ \\
     &    &   & (GeV) &   & (cm$^{-3}$) & & (G) \\ 
\tableline
1 &2.42$^{+0.09}_{-0.07}$& $=\alpha_p$ & [$8 \times 10^4$] &
   0.16$^{+0.29}_{-0.08}$ &  $>10^3$ & 0    &  $4.2 \times 10^{-5}$ \\
2 &2.25$^{+0.15}_{-0.25}$& $=\alpha_p$ & 25$^{+15}_{-10}$&
   0.50$^{+0.50}_{-0.40}$ & 4.0$^{+2.3}_{-2.7}$ & 1.07  & $6.6 \times 10^{-3}$\\
3 &2.35$^{+0.10}_{-0.20}$& 2.50$^{+0.20}_{-0.10}$ & [10$^3$] &
   0.13$^{+0.20}_{-0.03}$& 12.0$^{+7.}_{-4.}$ & 0.81 & $7.8 \times 10^{-5}$\\
4 &1.80$^{+0.10}_{-0.15}$& $=\alpha_p$ & 40$^{+23}_{-5}$&
   0.04$^{+0.06}_{-0.03}$& 0.06$^{+0.04}_{-0.03}$ & 5.57 & $1.7 \times 10^{-3}$
 \\
\tableline
\end{tabular}
\end{center}
\end{table*}

  Fit~2 ($\alpha_e = \alpha_p$, $E^{\rm max}_e$ fitted) is not
 very sensitive to the spectral index at acceleration, but is
 quite sensitive to the cutoff energy of the electrons. Fit
 prefers a low value of the electron cutoff energy $E_e^{\rm
 max}$, which eliminates the problem with the flat IC
 $\gamma$--ray spectrum and thus does not require very large
 matter density. Within 1$\sigma$ of the best fit values the
 matter density only varies by a factor of 2. Because of the low
 electron energy cutoff the fit allows for significantly flatter
 acceleration spectra.  The range of $R_e$ is significantly wider
 with $R_e$ decreasing from $\sim$1 for flat spectra to $\sim$0.1
 for spectral indices 1.35 -- 1.40.  The fit quality improves over Fit~1,
 especially for $\gamma$~Cygni. Adding one more parameter to the
 fit increases the error bars on all parameters that we
 evaluate. The best fit spectrum for IC 443 is shown in Fig.~6.

  Fit~3 ($\alpha_p, \; \alpha_e$ fitted, $E^{\rm max}_e$ fixed)
 also allows for a wide range of spectral indices. The most
 important restriction that data requires is an electron
 acceleration spectrum which is significantly steeper than the
 proton spectrum. Even these steep electron spectra do not fully
 compensate for the flat IC production spectrum, so that the
 matter density required is of order 1--10 cm$^{-3}$. The $R_e$
 value is quite stable, $\sim$0.1--0.2 or even less for IC
 443. The fit quality improves over Fit~1 by more than 1$\sigma$
 for both supernova remnants. Fig.~7 shows the best fit spectrum
 for IC 443 under the assumptions of Fit~3.

  Fit~4 ($\alpha_e=\alpha_p,\; E^{\rm max}_p = E^{\rm max}_e$
 fitted) produces the greatest improvement for both objects.
 This is especially true for $\gamma$~Cygni, for which all three
 previous sets of assumptions give significantly worse
 fits. Fig.~8 shows the best fit spectrum for $\gamma$~Cygni. It
 is important to notice that the contribution of the individual
 processes is now different. The role of bremsstrahlung is
 negligible. The $\gamma$--ray fluxes above 1 GeV are due to
 $\pi^0$ production while IC dominates at lower energy. The power
 law spectra preferred by the fits are very flat ($\alpha<$2),
 $R_e\;<$0.1, and the matter density is less than 0.1. Both
 electron and proton spectra cut off at less than 60 GeV.

 \section{Conclusions}

  With the current set of assumptions the fits of the
 $\gamma$--ray spectra of the two supernova remnants identified
 three possible sets of basic parameters. The first set,
 represented by Fit~1, consists of a relatively steep
 acceleration spectra and requires very large matter density in
 the acceleration region. The second set, Fits~2 \& 3 allows
 for  more reasonable matter densities at the $\gamma$--ray production
 region. Fit 4, which assumes a cutoff of the electron and proton
 spectra at the same maximum energy, results in a low matter density
 but also a very low value of the maximum energy.  A fit with $E_p^{max}$
 allowed to increase while keeping $E_e^{max}$ low would give a
 similar fit to the EGRET data, but steepening the electron spectrum
 through synchrotron radiation at such a low energy would require
 extreme magnetic fields. 

  Note that recent theoretical calculations, accounting for
 nonlinear effects (\cite{Berezhko96}; \cite{EllisonRey}),
 predict compression ratios higher than 4 that could lead to mean
 slopes at acceleration much flatter than the canonical $\alpha$
 = 2 value.  Flat electron spectra at low energies would also be
 more consistent with the radio spectral indices of the two
 supernova remnants which imply electron spectral indices at low
 energies of $\alpha_e = 1.72$ for IC443, and $\alpha_e = 2.0$
 for $\gamma$~Cyg (\cite{Green91}).  Fits 1 to 3 (IC~443) and Fits 1
 and 3 ($\gamma$~Cygni) appear to be inconsistent with the radio
 data.  However, if a high magnetic field were present
 (i.e. close to $H_\perp^{\rm max}$) then the energies of the
 electrons producing the radio synchrotron emission could be much
 lower than the energies of the $\gamma$-ray emitting electrons,
 thereby removing the inconsistency.  This would occur, for
 example in the case of Fit~2, where the synchrotron emitting
 electrons could have energies as low as 5~MeV (IC~443).

  The improvement of fit quality, shown in Tables 1 and 2,
 increases as expected when an additional fit parameter is
 introduced. We have not employed the $F$--test, which measures
 the fit improvement with the increase of number of parameters,
 because our different fits do not correspond to particular
 models of supernova remnants that we are testing. It is
 important to notice that $\ln{(L/L_0)}$ value is to a large
 extent determined by the end points of the data set where the
 efficiency and the statistical accuracy of the EGRET
 measurements are the worst.

 Following the EGRET observation of IC~443 and $\gamma$~Cygni it
 was anticipated, based on an $E^{-2}$ extrapolation of the
 observed flux, that these sources should be observable
 in the TeV range and above with ground based detectors.
 A simple $E^{-2}$ extrapolation of the Egret data
 is inconsistent with upper limits from CASA (\cite{Borione95})
 and Cygnus (\cite{Allen95}) around 100 TeV, from Hegra above 20 TeV
 and from Whipple above 300 GeV.  In this context, it is interesting
 to note that, except for Fit 4, which has very low $E^{max}$, our
 fits to the EGRET data alone seem to indicate steeper source
 spectra.

  The upper limits of the Whipple observatory on the TeV emission
 from the sources is shown on all emission spectra shown in
 Figs.~4 -- 8. It is important to notice that the fitted spectra
 are always below, or very close to, the experimental upper
 limits. In the case of Fit~1 of IC~443 (Fig.~4), for example,
 the fitted spectrum is $\sim$50\% higher than the upper limit.
 Half of that difference is due to the contribution of
 bremsstrahlung.  Note however, that if the interactions are
 taking place with neutral matter, such as that in a molecular
 cloud, then the cross sections we have used for bremsstrahlung
 in ionized matter would be inappropriate, and the bremsstrahlung
 spectral index could be steepened by as much as $\sim 0.1$ which
 would have the effect of bringing the fitted spectrum below
 the experimental limit.

  An electron source spectrum with $\alpha \sim 2.4$ is favored
 by a comparison of recent cosmic ray electron propagation
 calculations with direct cosmic ray electron observations above
 $\sim 10$ GeV
 (\cite{Strong94A,Strong95A,PorterProtheroe96}). However, source
 spectra as flat as $E^{-2}$ are possible (\cite{Aharonian95A})
 if the distance to the nearest source is $\sim 100$ pc or more.
 The spectra of the two SNR observed by EGRET show no sign of
 flattening at, or below, 100 MeV, indicating the presence of a
 steep spectral component due to bremsstrahlung.

 There are many different implications of a steep spectrum on
 acceleration ($\alpha \sim 2.2 - 2.4$), extracted with some of the
 assumptions of our analysis (Fits~1--3). For example, such a source
 spectrum for cosmic ray nuclei would require a loss rate
 in the galaxy, and hence a diffusion coefficient for cosmic ray
 propagation, proportional to $\sim E^{0.3}$ rather than $\sim
 E^{0.7}$ which is usually assumed based on cosmic ray secondary
 to primary ratios (note that a Kolmogorov spectrum actually
 predicts a diffusion coefficient proportional to $E^{1/3}$, see
 also Ptuskin et al. 1993, and Biermann 1995). Such a diffusion
 coefficient is also favored by reacceleration models
 (\cite{HeinbachSimon95,SeoPtuskin94}). Such spectra, however, would
 not be consistent with the radio observations and would require
 a strong spectral steepening above radio emitting energies.

 In conclusion, we have modeled the gamma ray emission of IC~443
 and $\gamma$~Cygni including inverse Compton scattering on all
 relevant radiation fields, bremsstrahlung, and pion
 production. We fit the modeled production spectra to the EGRET
 data assuming a power-law momentum spectrum of electrons and
 nuclei in the SNR and exponential cutoffs. The results from the
 fitting show that (a) the dominant contributions come from
 bremsstrahlung and $\pi^0$ decay provided the spectra of both
 electrons and nuclei extend above 60 GeV. The IC contribution
 can only be important with fine tuning, when $E^{\rm max}$ for both
 electrons and nucleons is $\sim$40 GeV; (b) an electron to
 proton ratio of $\sim$ 0.05--0.5 is required under all four
 assumptions that we have explored; (c) a spectral index of
 $\alpha \sim 2.2$--2.4 is required if electrons and nuclei are
 accelerated with the same spectral indices unless the
 acceleration spectra cut off below 60 GeV; (d) if the
 acceleration spectra of electrons and nuclei are not the same
 the data requires electron spectra that are significantly
 steeper than the nuclei spectra, which contribute to the
 experimentally observed range only the $\pi^0$ feature, which
 depends mildly on the acceleration spectra of nuclei; (e) a very
 high, but uncertain, average matter density is required if the
 electron spectrum has the same spectral index as the proton
 spectrum and the cut-off energy is very high, e.g. 80 TeV (note,
 however, that matter densities as high as $\sim 300$ cm$^{-3}$
 are present in molecular clouds associated with these SNR).  We
 also note that in this case the bremsstrahlung contribution at 1
 GeV is $\sim 40\%$, somewhat higher than the $\sim 10\%$
 obtained from a fit to the diffuse galactic gamma ray intensity
 (\cite{Hunter95,Hunter97}), which is probably lower due to the effects of
 cosmic ray energy losses during propagation which are different
 for electrons and nuclei.

  At the present time and with the current available experimental 
 information we cannot make a choice between the two basic assumptions
 that we used for fitting the EGRET data. Additional experimental
 information is necessary for further analysis. One crucial data set
 would consist of TeV detections (or lower upper limits) of these
 two and other supernova remnants. A possible observation would allow
 us the exclude the assumption for a low energy cutoff of the acceleration
 spectra and to obtain better determined values for the shape of the
 electron and proton spectra and for the matter density at the source.
 We also plan to extend our analysis by using theoretically motivated 
 shapes for the spectra of the accelerated particles. It will then be
 possible to also predict the expected 100 TeV $\gamma$--ray fluxes
 from supernova remnants in some current models (\cite{Berezhko96}),
 which extend the accelerated particles spectra to Z$\times$400 TeV.\\

 \noindent {\bf Acknowledgements.} We thank Joe Esposito and the
 EGRET group for providing the measured spectra of IC~443 and
 $\gamma$~Cygni in numerical form. We are grateful to the referee
 Steven Reynolds for helping us to improve on the first version
 of this paper. RJP thanks the Bartol Research Institute for
 hospitality during part of 1996. We thank Troy Porter for a
 useful discussion and for reading the manuscript. The research
 of TKG is supported in part by NASA Grant NAGW--4605. TS is
 supported in part by NASA grant NAGW--3880. The research of RJP
 is supported by a grant from the Australian Research Council.

\clearpage

\begin{figure}
\vspace{16cm}
\includegraphics{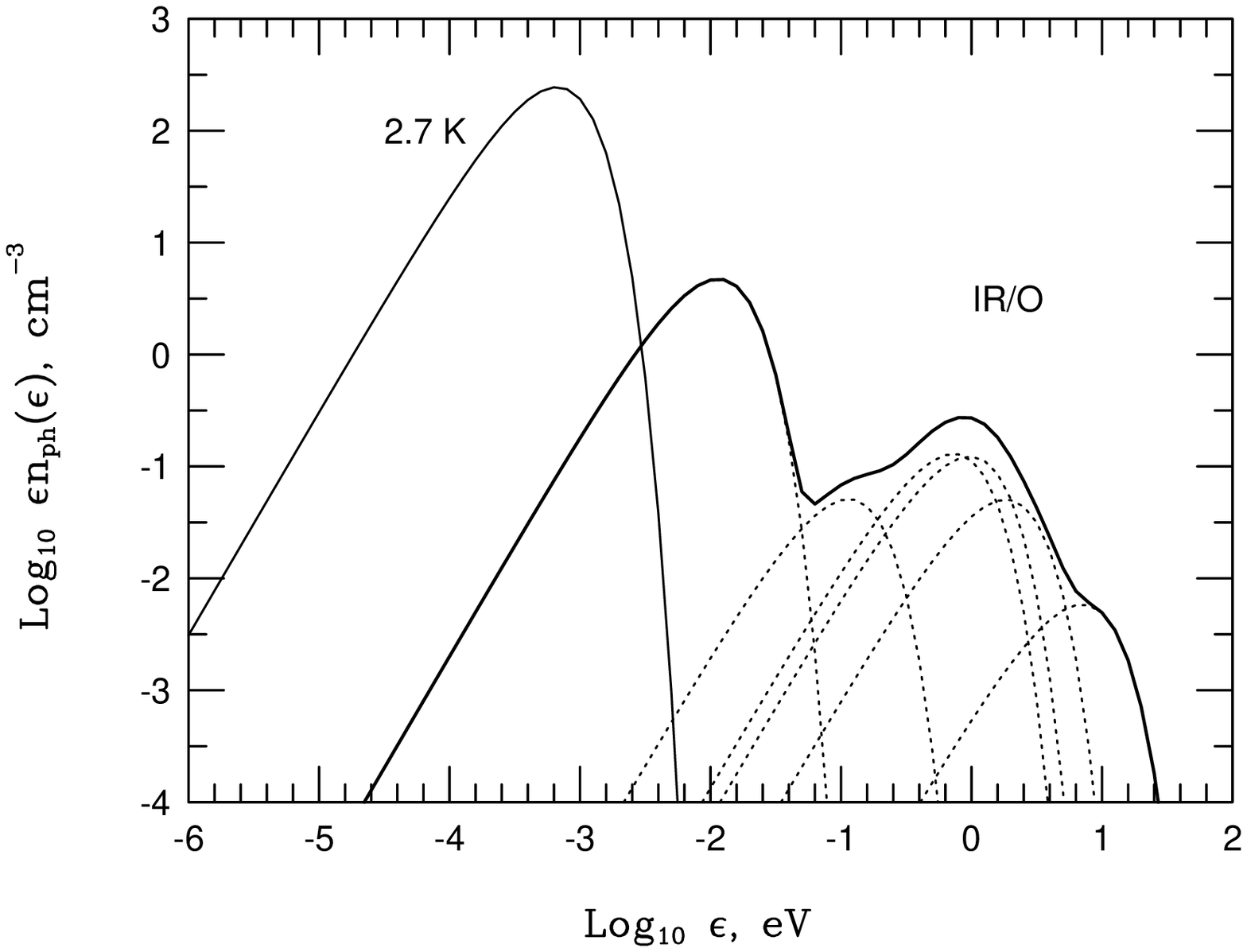}
\caption{The interstellar infrared/optical (IR/O) radiation
field based on the work of Mathis, Mezger, \& Panangia (1983)
composed of six diluted blackbody spectra (dotted lines), and the
cosmic microwave background (2.7 K).}
\label{fig:fig1}
\end{figure}

\setcounter{figure}{1}
\begin{figure}
\vspace{16cm}
\includegraphics{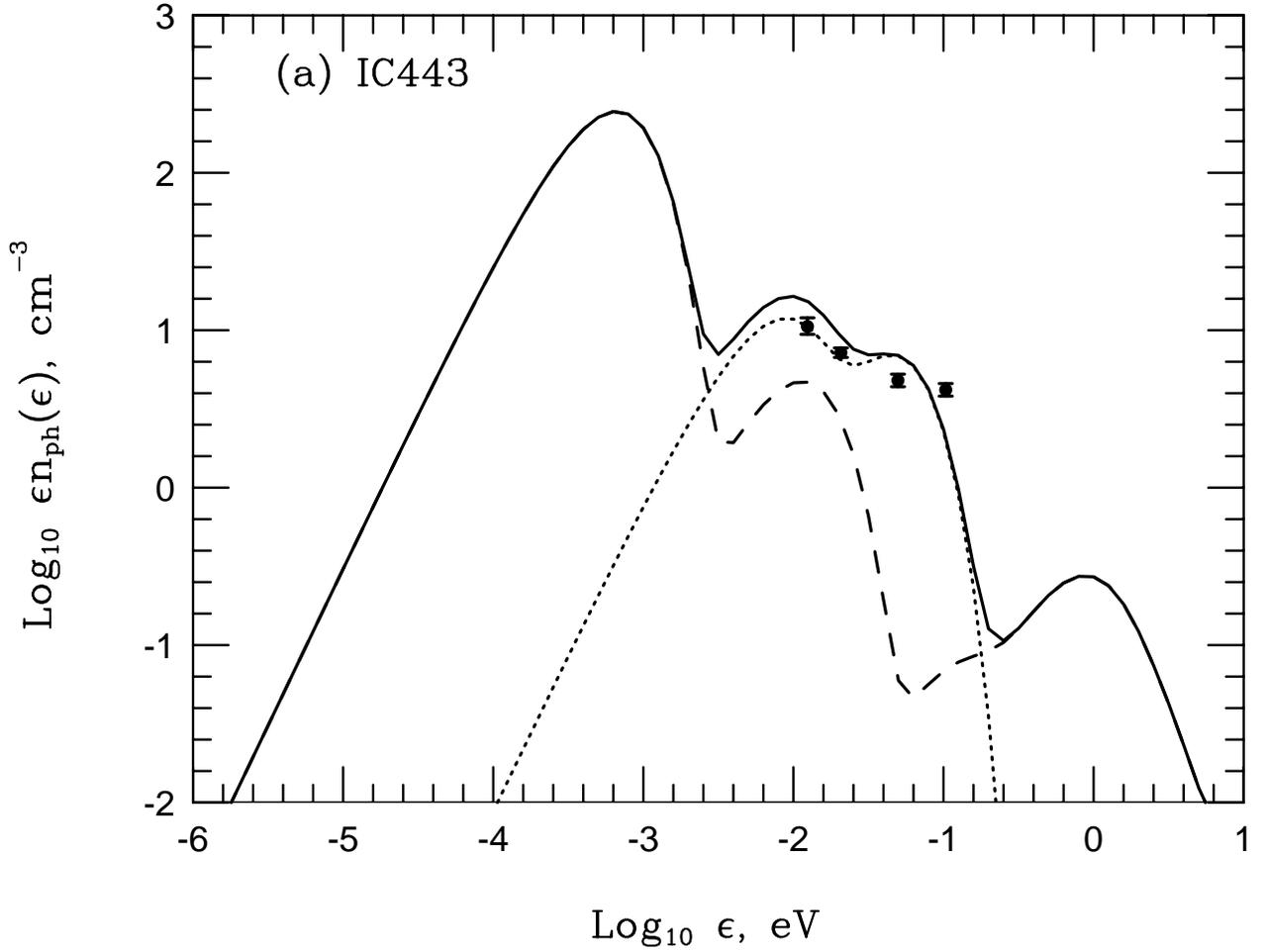}
\caption{Total radiation field at the supernova remnant (solid
curve) showing contributions from interstellar infrared/optical
plus microwave (dashed line) and two-temperature fit to infrared
spectra (\protect \cite{Saken92}) (dotted curve) and recent data
for (a) IC~443 (data from \protect \cite{Mufson86}) and (b)
$\gamma$~Cygni (data from \protect \cite{Saken92}).}
\label{fig:fig2}
\end{figure}

\setcounter{figure}{1}
\begin{figure}
\vspace{16cm}
\includegraphics{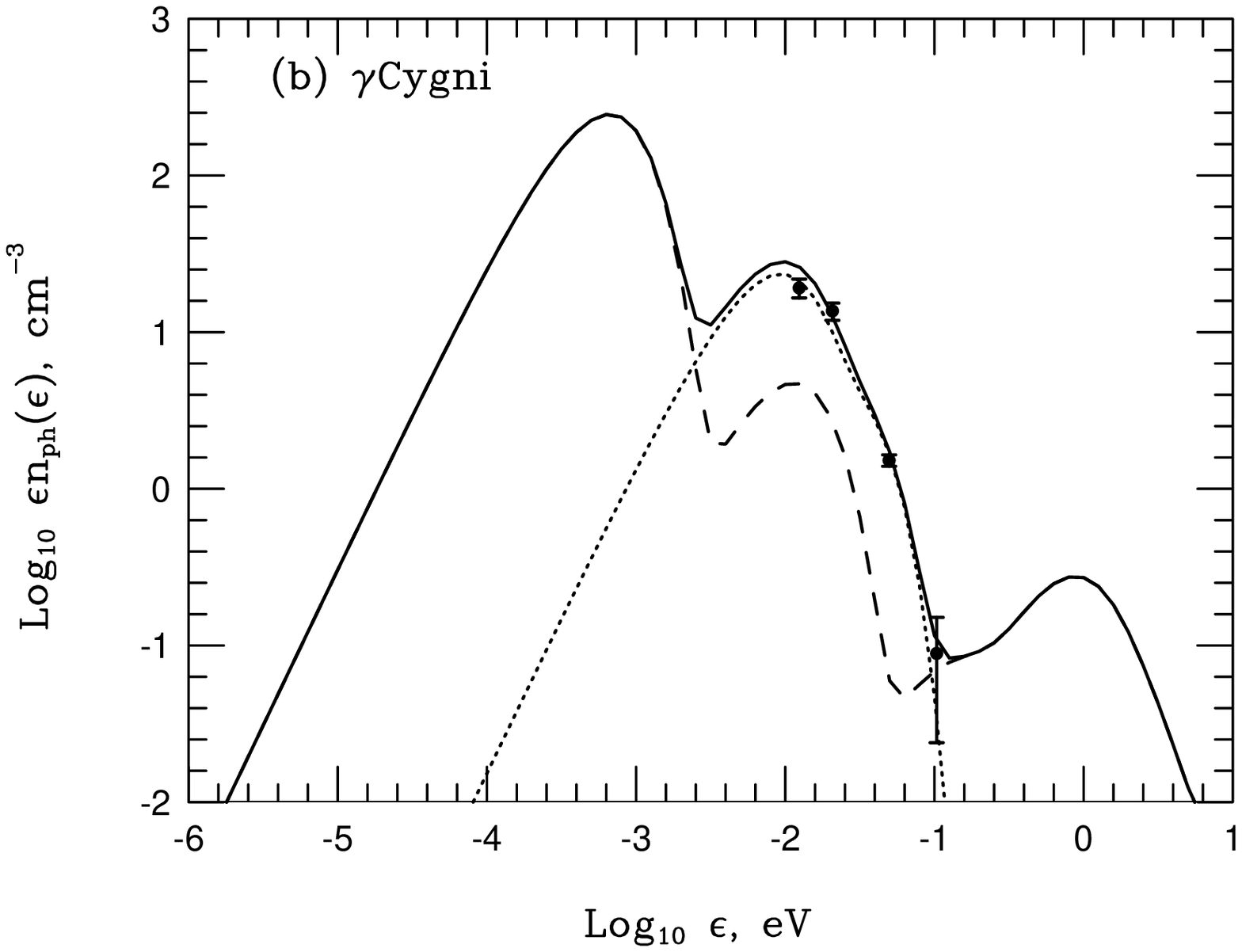}
\caption{(b) }
\end{figure}

\begin{figure}
\vspace{16cm}
\includegraphics{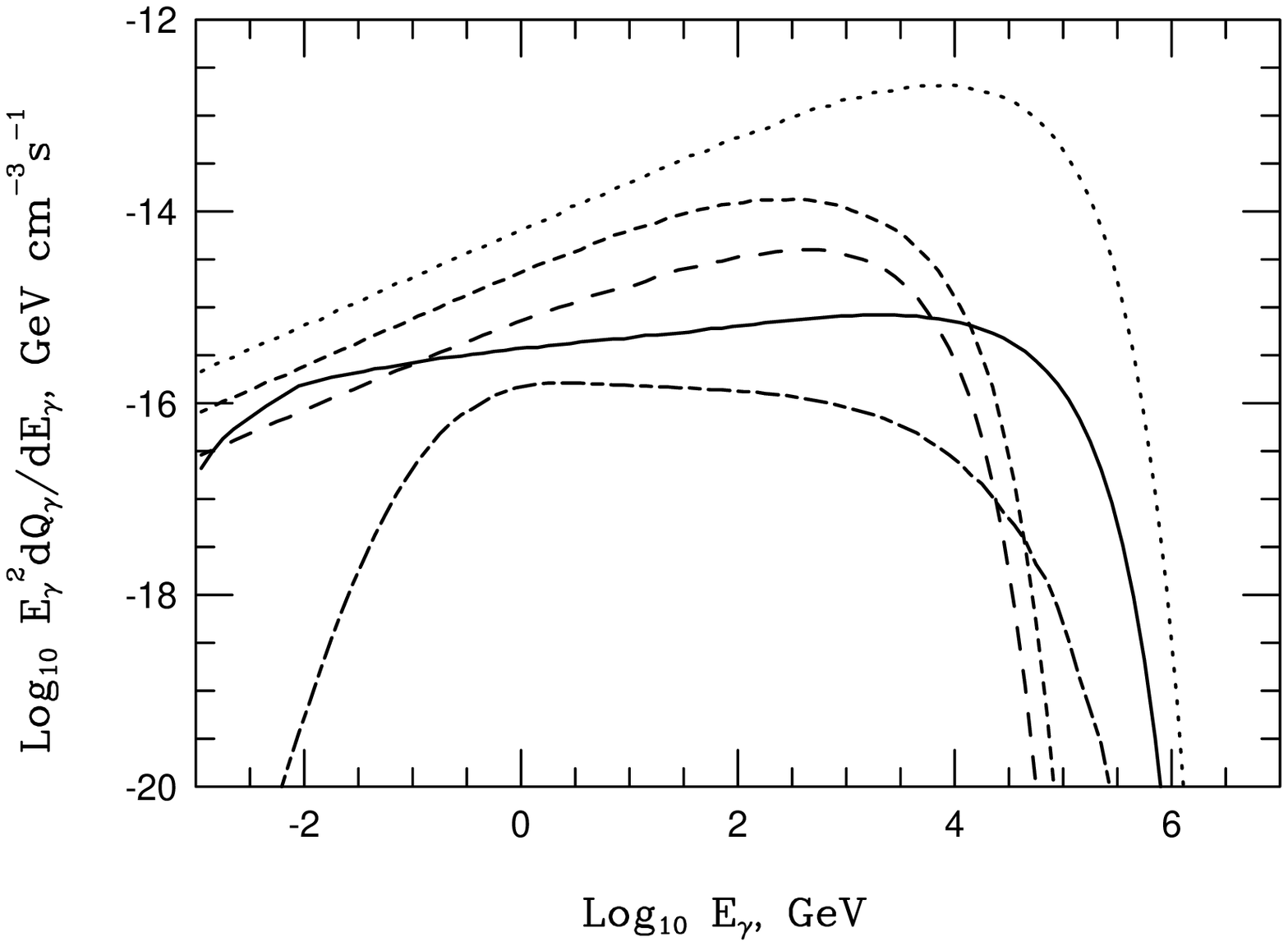}
\caption{Gamma ray emissivity at IC~443 produced by particles
(electrons and protons) with momentum spectrum $dn/dp = p^{-2}$
cm$^{-3}$ GeV/c$^{-1}$ interacting with interstellar matter with
nucleon number density 1 cm$^{-3}$ and with interstellar (IR/O)
and cosmic microwave radiation fields of Figure~1: dot-dash curve
-- $\pi^0$ production; solid curve -- bremsstrahlung; long dashes
-- inverse Compton on interstellar (IR/O); short dashes --
inverse Compton on Infrared radiation at IC~443; dotted curve --
inverse Compton on microwave.}
\label{fig:fig3}
\end{figure}

\begin{figure}
\vspace{16cm}
\includegraphics{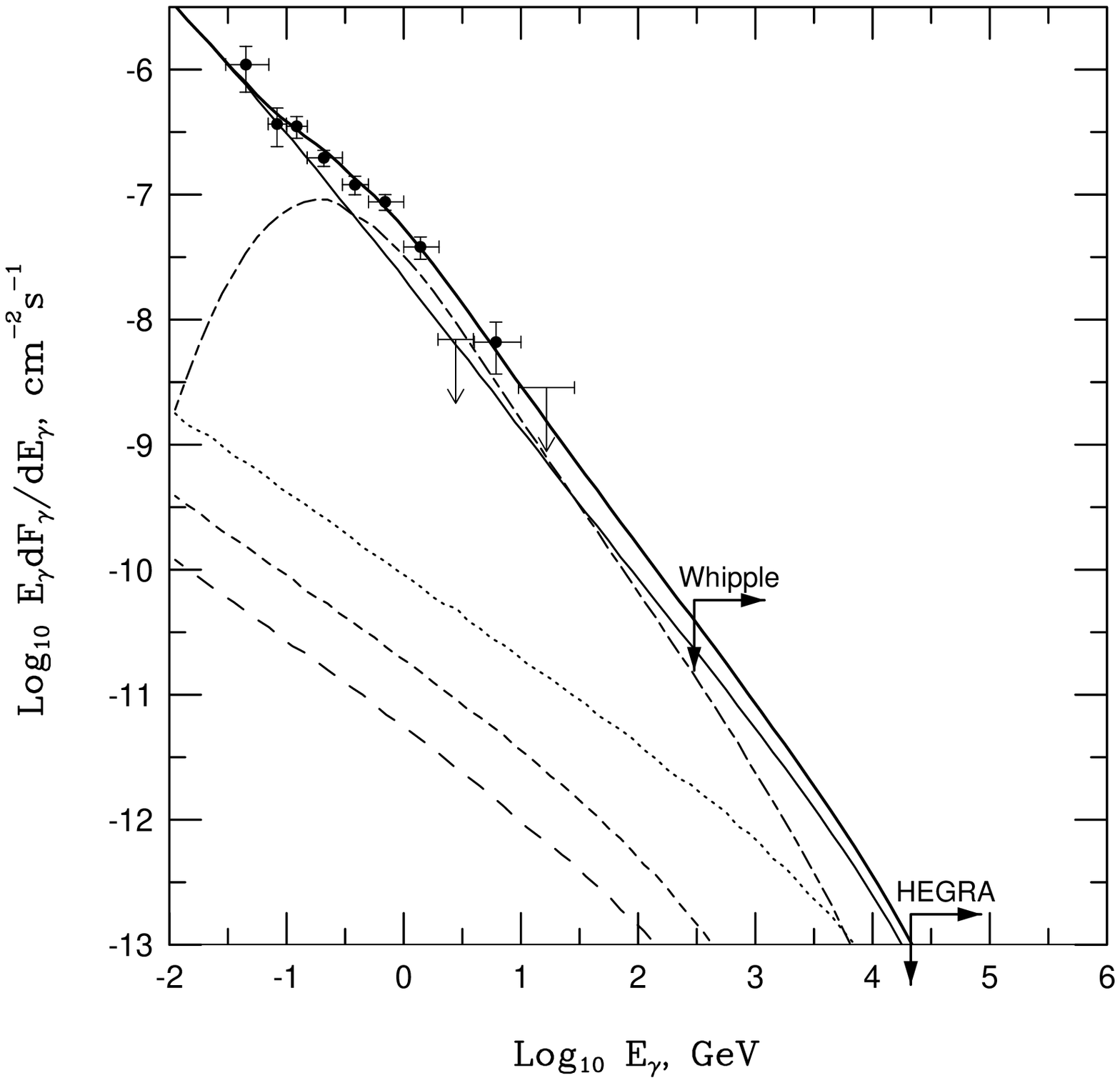}
\caption{Best fit (Fit~1) to EGRET observations of IC~443,
 including upper limits (thick solid curve). Whipple (\protect
 \cite{Buckley97}) and HEGRA upper limits (\protect
 \cite{Prosch95}) are not included in the fits.  The small IC
 contributions from the interstellar and source IR radiation are
 not shown in this and other fit spectra.  The other curves have
 the same meaning as in Fig.~3.}
\label{fig:fig4}
\end{figure}

\begin{figure}
\vspace{16cm}
\includegraphics{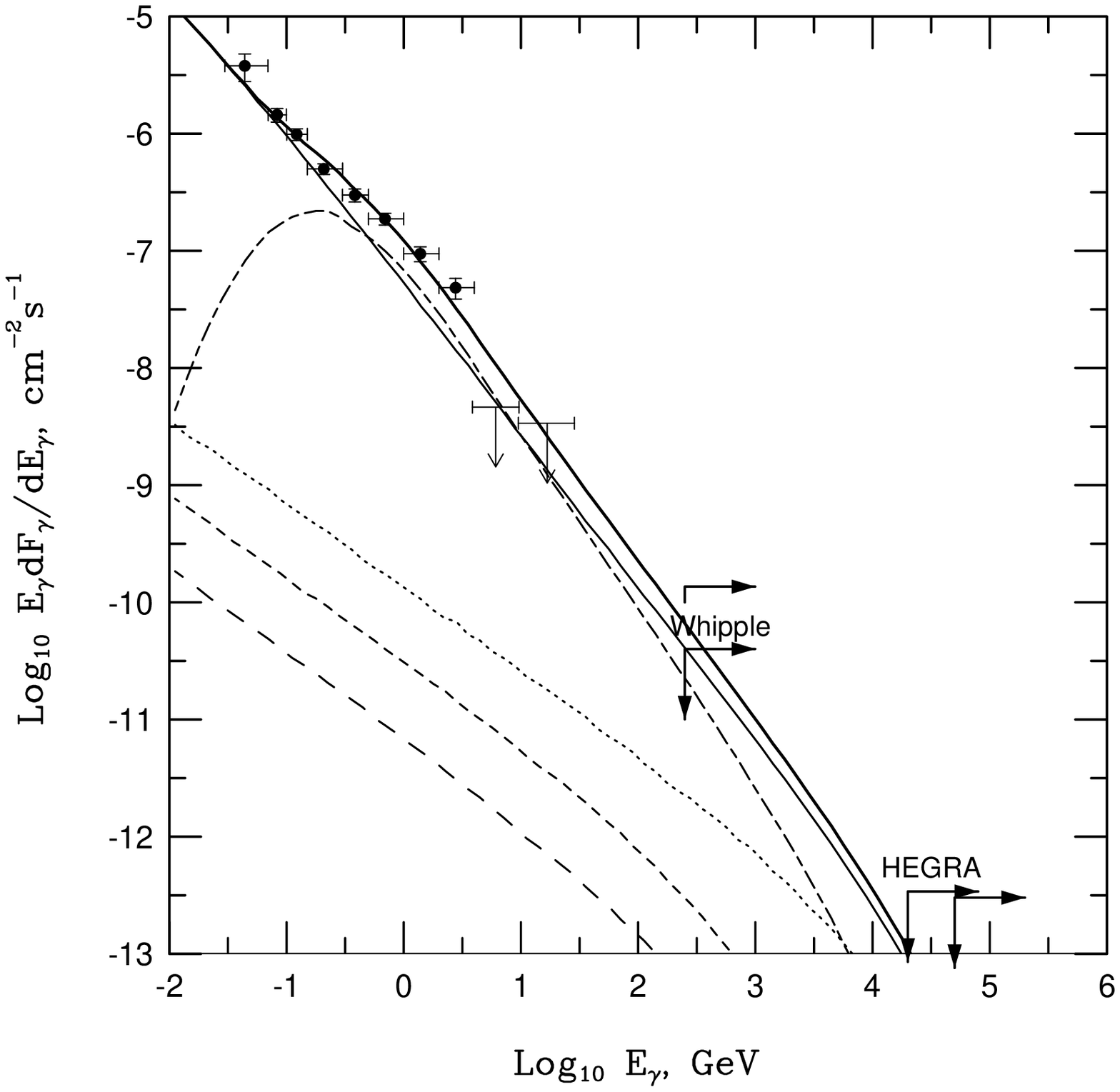}
\caption{Best fit (Fit~1) to EGRET observations of
 $\gamma$~Cygni, including upper limits (thick solid
 curve). Whipple (\protect \cite{Buckley97}) and HEGRA upper
 limits (\protect \cite{Prosch96,Willmer95}) are not included in
 the fits.}
\label{fig:fig5}
\end{figure}

\begin{figure}
\vspace{16cm}
\includegraphics{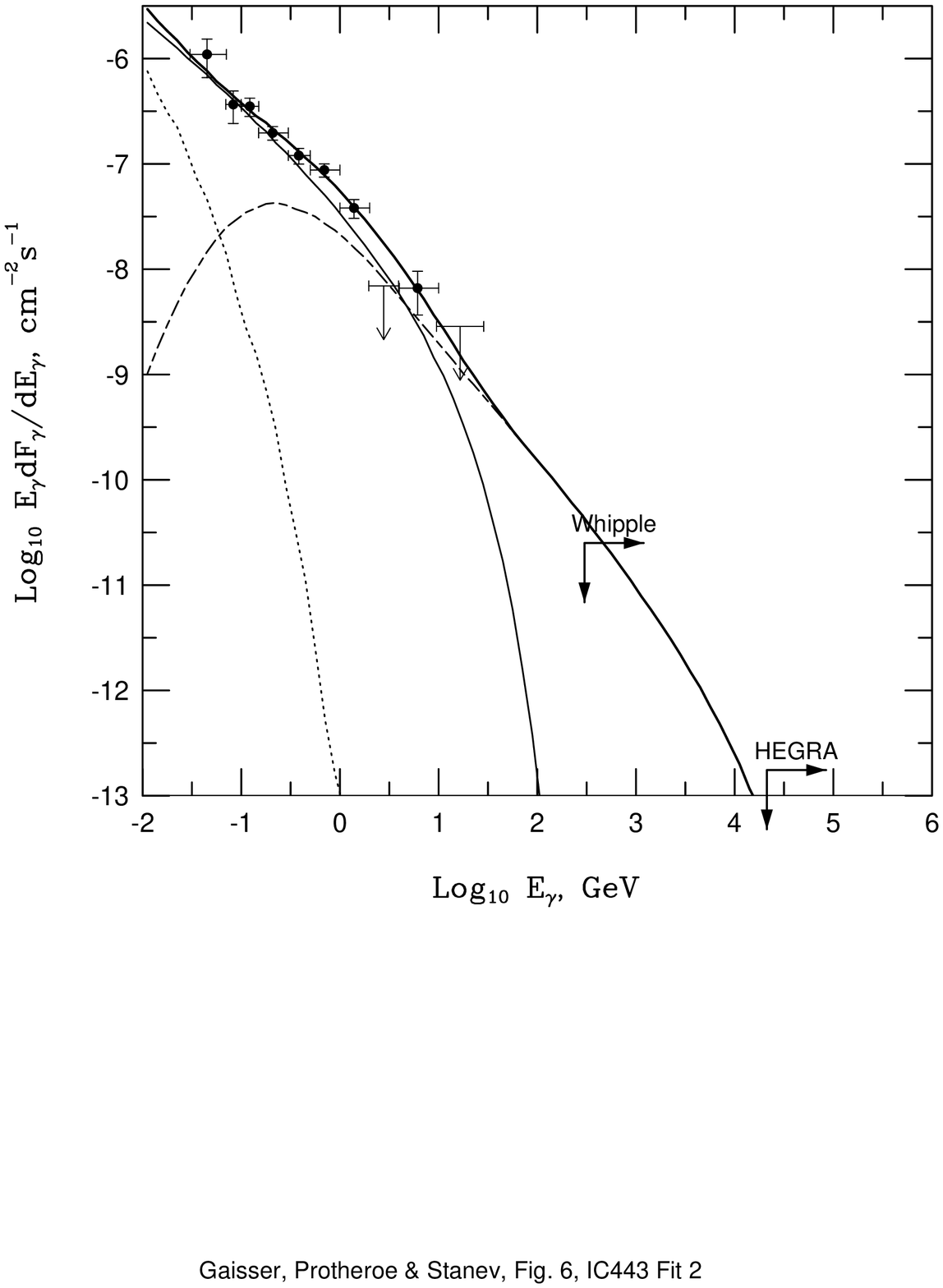}
\caption{Best fit (Fit~2) of IC~443.}
\label{fig:fig6}
\end{figure}

\begin{figure}
\vspace{16cm}
\includegraphics{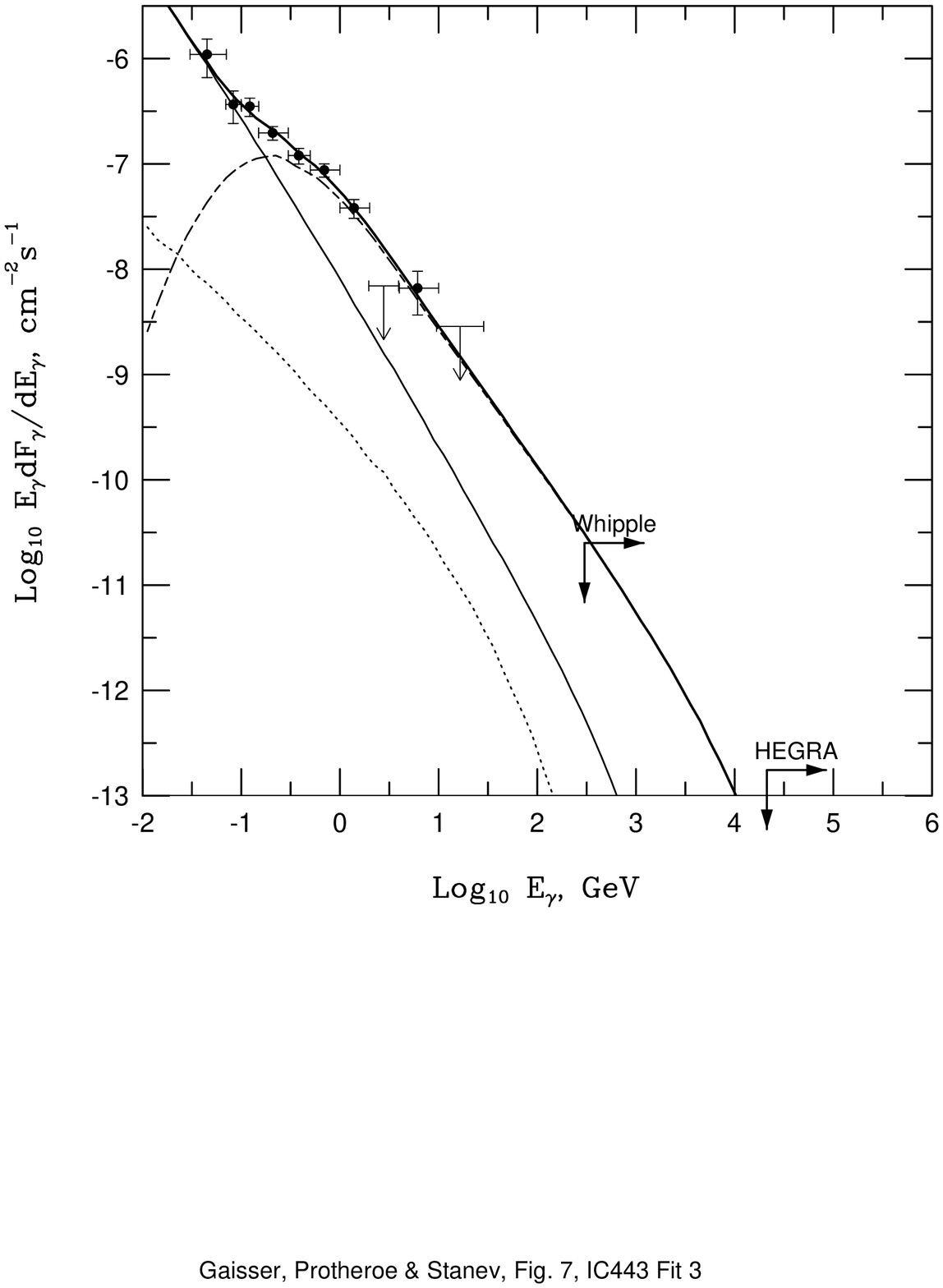}
\caption{Best fit (Fit~3) of IC~443.}
\label{fig:fig7}
\end{figure}

\begin{figure}
\vspace{16cm}
\includegraphics{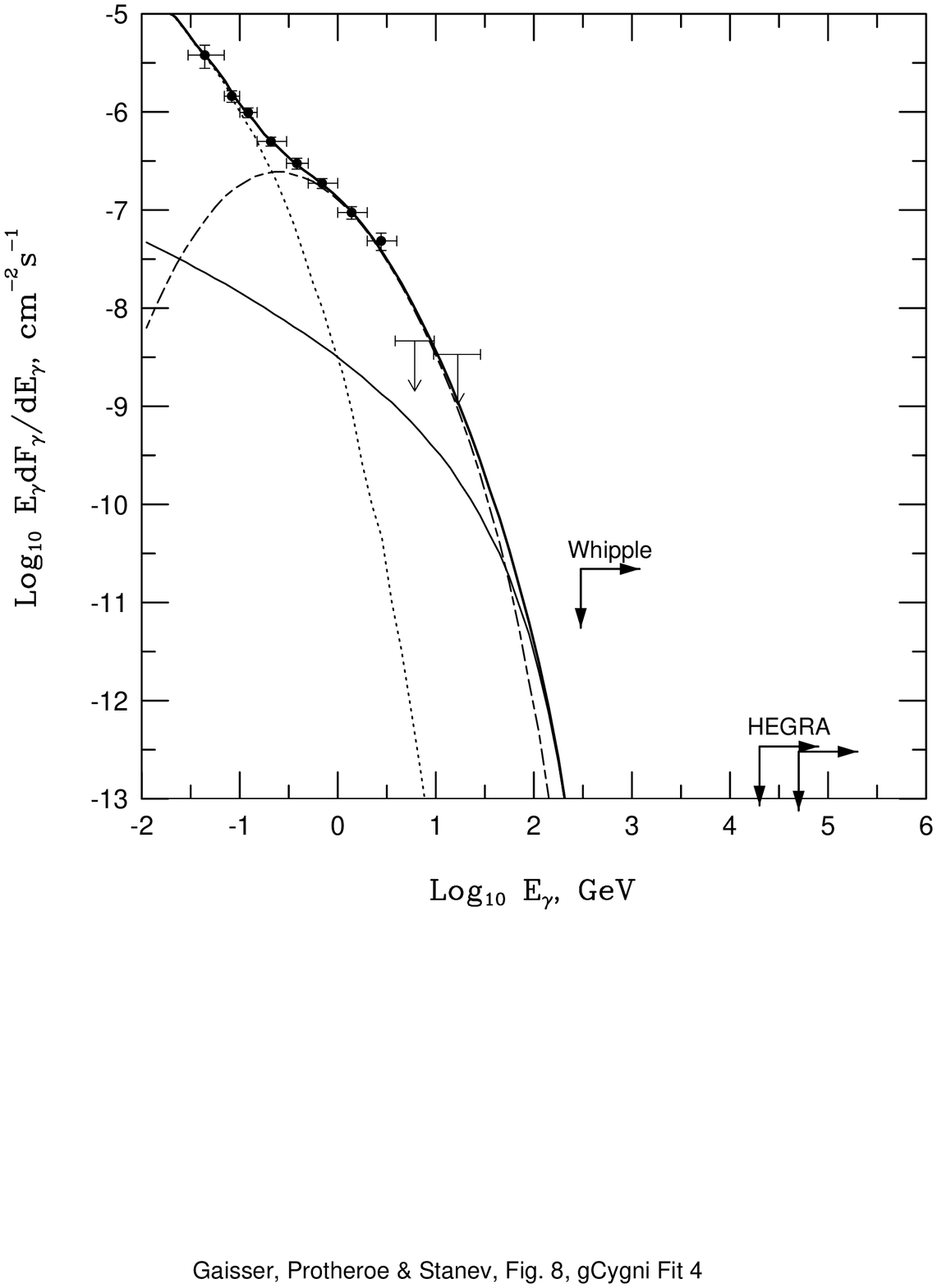}
\caption{Best fit (Fit~4) of $\gamma$~Cygni.}
\label{fig:fig8}
\end{figure}

\end{document}